\documentclass{article}
\usepackage{amsmath,graphicx,amssymb,bm,units,subfigure,acronym,color,enumerate,balance,graphics,graphicx,psfrag,blindtext}
\usepackage[english]{babel}
\usepackage[sort]{cite}
\usepackage[export]{adjustbox}
\usepackage{waspaa17,amssymb,amsmath,graphicx,times,url}
\usepackage{color}

\definecolor{myred}{rgb}{0.9,0.1,0.1}

\title{Broadband DOA estimation using convolutional neural networks trained with noise signals}

\name{Soumitro Chakrabarty and Emanu\"{e}l A.~P.~Habets}
\address{International Audio Laboratories Erlangen\sthanks{\,A joint institution of the Friedrich-Alexander-University Erlangen-N\"urnberg (FAU) and Fraunhofer Institute for Integrated Circuits (IIS).},  
Am Wolfsmantel 33, 91058 Erlangen, Germany\\
\{soumitro.chakrabarty, emanuel.habets\}@audiolabs-erlangen.de}

\begin{document}

\ninept
\maketitle
\setlength{\textfloatsep}{7pt}
\setlength{\floatsep}{5pt}
\begin{sloppy}

\begin{abstract}
A convolution neural network (CNN) based classification method for broadband DOA estimation is proposed, where the phase component of the short-time Fourier transform coefficients of the received microphone signals are directly fed into the CNN and the features required for DOA estimation are learned during training. Since only the phase component of the input is used, the CNN can be trained with synthesized noise signals, thereby making the preparation of the training data set easier compared to using speech signals. Through experimental evaluation, the ability of the proposed noise trained CNN framework to generalize to speech sources is demonstrated. In addition, the robustness of the system to noise, small perturbations in microphone positions, as well as its ability to adapt to different acoustic conditions is investigated using experiments with simulated and real data.  
\end{abstract}


\section{Introduction}
\label{sec:intro}
Many applications such as hands-free communication, teleconferencing, and distant speech recognition require information on the location of a sound source in the acoustic environment. The relative location of a sound source with respect to a microphone array is generally given in terms of the direction of arrival (DOA) of the sound wave originating from that location.  In most practical scenarios, this information is not available and the DOA of the sound source needs to be estimated. However, accurate DOA estimation is a challenging task in the presence of noise and reverberation. 

Over the years, several methods have been developed for the task of broadband DOA estimation. Some popular approaches are: \emph{i)} subspace based approaches such as multiple signal classification (MUSIC) \cite{Schmidt1986}, \emph{ii)} time difference of arrival (TDOA) based approaches that use the family of generalized cross correlation (GCC) methods \cite{Knapp1976, Huang2001c}, \emph{iii)} generalizations of the cross-correlation methods such as steered response power with phase transform (SRP-PHAT) \cite{Brandstein1997}, and multichannel cross correlation coefficient (MCCC) \cite{Benesty2008a}, and \emph{iv)} model based methods such as maximum likelihood method \cite{Stoica1990}. These traditional methods generally suffer from problems such as high computational cost and/or degradation in performance in presence of noise and reverberation\cite{Benesty2008a}. 

Recently, deep neural networks (DNN) based supervised learning methods have shown success in various fields ranging from computer vision \cite{Krizhevsky2012} to speech recognition \cite{Hinton2012}. Following this, different DNN based methods have been proposed for the task of DOA estimation \cite{Takeda2016, Xiao2015, Takeda2016a}. These methods generally involve an explicit feature extraction step. While in \cite{Xiao2015} GCC vectors are provided as input to the learning framework, in \cite{Takeda2016, Takeda2016a} the eigenvalue decomposition of the spatial correlation matrix is performed to provide the eigenvectors corresponding to the noise subspace as input. Along with the extra computational cost involved in the feature extraction, these methods can potentially suffer from the same problems as the traditional methods. 

In this paper, we propose a convolution neural network (CNN) based classification method for broadband DOA estimation. CNNs are a variant of the standard feed-forward network that compute neuron activations through shared weights over small local areas of the input \cite{Krizhevsky2012}. Rather than involving an explicit feature extraction step, the phase component of short-time Fourier transform (STFT) coefficients of the input signal is directly provided as input to the neural network, and the CNN learns the information required for DOA estimation during training. Using only the phase information also makes it possible to train the system with synthesized noise signals rather than real-world signals like speech. This makes the preparation of the training data set easier. Through experimental evaluation, we investigate the ability of the noise signal trained system to generalize to speech sources as well as the robustness of the system to noise and small perturbations in the microphone positions. We also investigate the ability of the proposed system to adapt to different acoustic conditions.     

\section{DOA estimation as a classification problem}
\label{sec:Prob}

In this work, we want to utilize a CNN based framework for DOA estimation, where the aim is to learn a mapping from the observed microphone array signals to the DOA of the impinging sound wave using a large set of labeled training data. The DOA estimation is performed for each time frame of the short-time Fourier transform (STFT) representation of the observed signals.

The problem of DOA estimation is formulated as an $I$-class classification problem, where each class corresponds to a possible DOA value in the set $\Theta = \lbrace \theta_{1}, \ldots, \theta_{I} \rbrace$, and the DOA estimate is given as the DOA class with the highest posterior probability. The number of classes, $I$, depends on the array geometry as well as the resolution for discretization of the whole range of DOAs. For example, for a uniform linear array (ULA) the DOA range lies between $[0^{\circ},180^{\circ}]$, and with a resolution of $2^{\circ}$, the total number of classes is $I = 91$. 

A supervised learning framework comprises of a training and a test phase. In the training phase, the DOA classifier is trained on a training data set, consisting of pairs of fixed dimension feature vectors and their corresponding DOA class labels. In the test phase, given an input feature vector, the classification system generates the posterior probability for each of the $I$ DOA classes based on which the DOA estimate is obtained.

\section{CNN based DOA estimation}
\label{sec:CNNDOA}
In this section, we first describe the specific input feature representation used in this work followed by details regarding CNN and its application to DOA estimation.  
\subsection{Input feature representation}
\label{ssec:IP}
The first challenge is to find a feature representation that contains sufficient information for DOA estimation. As a first step, the received microphone signals are transformed to the STFT domain using an $N_f$ point discrete Fourier transform (DFT). Note that in the STFT domain the observed signals at each TF instance are represented by complex numbers. Therefore, the observed signal can be expressed as
\begin{equation}
Y_{m}(n,k) = A_{m}(n,k) e^{j\phi_{m}(n,k)} ,
\end{equation}
where $A_{m}(n,k)$ represents the magnitude component and $\phi_{m}(n,k)$ denotes the phase component of the STFT coefficient of the received signal at the $m$-th microphone for the $n$-th time frame and $k$-th frequency bin.

In this work, rather than having an explicit feature extraction step, we directly provide the phase component of the STFT coefficients of the received signals as input to our system. The idea is to make the system learn the relevant feature for DOA estimation from the phase component through training. 

Since the aim is to compute the posterior probabilities of the DOA classes at each time frame, the input feature for the $n$-th time frame is formed by arranging $\phi_{m}(n,k)$ for each time-frequency bin $(n,k)$ and each microphone $m$ into a matrix of size $M \times K$, which we call the \emph{phase map}, where $K = N_{f}/2 +1$ is the total number of frequency bins, upto the Nyquist frequency, at each time frame and $M$ is the total number of microphones in the array. For example, if we consider a microphone array with $M=4$ microphones and $N_{f} = 256$, then the input feature matrix is of size $4 \times 129$. Given the input representations, the next task is to estimate the posterior probabilities of the $I$ DOA classes. For this, we propose a CNN based supervised learning method, described in the following subsections.

\subsection{Convolutional neural networks - Basics}
\label{ssec:CNN}
\begin{figure}[t]
\centering
\includegraphics[scale=0.3]{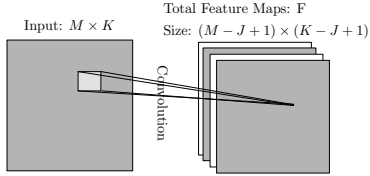}
\caption{Illustrative diagram to show the convolution operation in convolution layers of CNN. We consider F different local filters each of size $J \times J$.}
\label{fig:Conv}
\end{figure}
CNNs are a variant of the standard fully-connected neural network, where the architecture generally consists of one or more ``convolution layers'' followed by fully-connected layers leading to the output. In typical CNN architectures, the convolution layers are pairs of convolution and pooling operation. In the convolution operation, a set of filters is applied that process small local parts of the input. The individual elements of these filters are the weight parameters that are learned during training and the application of each filter to the input generates a feature map at the output. 

An illustration of the convolution operation is shown in Figure~\ref{fig:Conv}. In the illustration, we consider local filters of size $J \times J$ and a 2D convolution is performed by moving the filter across both dimensions of the input of size $M \times K$ in steps of 1 element to generate feature maps of size $(M-J+1) \times (K-J+1)$. Here, we consider $F$ different filters, that results in $F$ feature maps following the convolution operation. As each filter is applied across the whole input space, it leads to a critical concept in CNNs, called ``weight sharing", which leads to fewer trainable parameters compared to fully connected networks \cite{AbdelHamid2012}. 

The convolution operation is then followed by an activation layer, which operates point-wise over each element of the feature maps at the output of a convolution operation. This is followed by pooling where the aim is to reduce the feature map resolution by combining the filter activations from different positions within a specified region. Finally, the fully connected layers aggregate information from all different positions to perform the classification of the complete input. For further details on CNNs, the reader is referred to \cite{LeCun1998}. 
\subsection{DOA estimation with CNNs}
\label{ssec:CNNest}
With the phase map as the input, the task of the CNN is to generate the posterior probabilities for each of the DOA classes. Let us denote the phase map for the $n$-th time frame as $\bm{\Phi}_{n}$. Then the posterior probability generated by the CNN at the output is given by $p(\theta_i|\bm{\Phi}_n)$, where $\theta_i$ is the DOA corresponding to the $i$-th class. In Figure~\ref{fig:Arch}, we show the CNN architecture employed in this work. In the convolution layers (Conv layers in Figure~\ref{fig:Arch}), small filters of size $2\times2$ are applied to learn local correlations between the phase components of neighboring microphones at local frequency regions. These learned local structures are then eventually combined by the fully connected layers (FC layers in Figure~\ref{fig:Arch} ) for the final classification task. 

Applying local filters can potentially lead to better robustness against noise \cite{AbdelHamid2012}. In the presence of noise, the signal-to-noise ratio (SNR) across the spectrum is not constant, therefore the filters can detect local phase structures from the high SNR part well enough to compensate for the lack of information from the low SNR regions. Due to the weight sharing concept in CNNs, they also provide robustness to local distortions in the input \cite{LeCun1998}. Therefore, applying the filters to learn local phase structure over neighboring microphones can provide additional robustness to small perturbations in microphone positions.  

For both the convolution as well as the fully connected layers, in this work, we use the rectified linear units (ReLU) activation function \cite{Nair2010}. In contrast to conventional CNN architectures, we do not have any pooling layer. In our experiments, inclusion of pooling layers showed a slight decrease in performance. 

In the final layer of the network, we use  the softmax activation function to perform classification. The softmax function generates the posterior probability for each of the $I$ classes. Given the posterior probabilities, the final DOA estimate is given by 
\begin{equation} \label{eq:EST}
\hat{\theta}_{n} = \operatorname*{arg \; max}_{\theta_{i}} \; p(\theta_{i}|\bm{\Phi}_n). 
\end{equation}

The number of convolution layers, fully connected layers and the network parameters in the proposed architecture in Figure~\ref{fig:Arch} was chosen by using a validation data set. Through various experiments with different sized networks, the architecture with the minimum average validation loss over data from different acoustic conditions was chosen as the final architecture.  

The CNN is trained using a training data set $\lbrace \lbrace \bm{\Phi}_n, \theta_n \rbrace \;  | \; n = 1,\ldots, N \rbrace$, where $N$ denotes the total number of STFT time frames in the training set. Details regarding the preparation of the training data set are given in Section~\ref{ssec:Training}.

In the test phase, the test signals are first transformed into the STFT domain using the same parameters used during training. Following this, the phase map for each time frame of the test signals is given as input to the CNN, and the CNN generates the posterior probabilities of the $I$ DOA classes. The final DOA estimate for each time frame of the test signals is given by (\ref{eq:EST}).  
\begin{figure}[t]
\centering
\includegraphics[width = 0.47\textwidth]{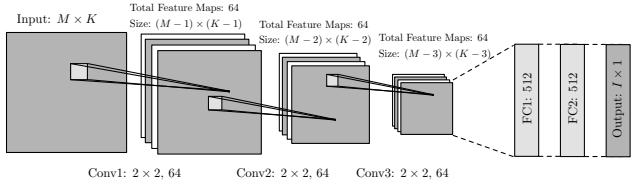}
\caption{Proposed CNN architecture.}
\label{fig:Arch}
\end{figure}
\section{Training with noise}
\label{sec:training}
As mentioned earlier, our input feature representation consists of only the phase part of the STFT coefficients of the signal. Since the magnitude spectrum is not utilized, it is possible to prepare the training data set using synthesized signals rather than using actual speech recordings. In this work, we train the proposed neural network using spectrally white noise sources positioned at different angles and distances relative to the microphone array.  

There are some significant advantages of being able to train the network with noise signals. First, for preparation of the training data set, we do not require any speech databases. Second, it makes the design of ground truth labels easier. When using speech signals, a voice activity detector (VAD) is generally required to detect silent frames \cite{Xiao2015, Takeda2016}, since features from silent frames do not contain useful patterns for training. Errors in detecting silent frames can lead to inconsistent labels leading to error in training. Such problems can be avoided when using synthesized noise signals for training.     

\section{Experimental results}
\label{sec:EXP}
In this section, we present the experimental evaluation results, where the performance of the proposed method is compared to the traditional broadband DOA estimation method, SRP-PHAT \cite{Brandstein1997}. Since we propose a classification approach to DOA estimation, similar to \cite{Takeda2016}, the performance is evaluated in terms of frame level accuracy, which can be given by 
\begin{equation}
A (\%) = \frac{\hat{N}^{}_{c}}{N_s} \times 100,
\end{equation}
where $N_{s}$ denotes the total number of time frames in the test data set where speech is active and $\hat{N}_{c}$ denotes the number of such time frames where the estimated DOA corresponds to the true DOA. Since we have access to the clean speech signals, the time frames containing speech can be easily determined.
\begin{table}[t]
\footnotesize
\centering
    \setlength{\tabcolsep}{3 mm}
       \vspace{1 em}
	\begin{tabular}{c | c }
		\hline \hline
		   \multicolumn{2}{c}{{Simulated training data}}	\\
		 \hline 
		  Signal & Synthesized noise signals \\
		  \hline
		  Room size  & R1: ($6 \times 6$) m , R2: ($5 \times 5$) m \\
		  \hline
		  Array positions in room & 7 different positions in each room \\
		  \hline
		  Source-array distance& 1 m and 2 m  for each position\\
		  \hline
		  RT$_{60}$ & R1: 0.3 s, R2: 0.2 s \\
		  \hline
		  SNR & Uniformly sampled from 0 to 20 dB \\		
		  \hline \hline
	\end{tabular}
	\caption{Configuration for training data generation. All rooms are 2.5 m high.}
\label{tab:Train}\vspace{0.1em}
\end{table}
\vspace{-0em}
\begin{table}[t]
\footnotesize
\centering
    \setlength{\tabcolsep}{3 mm}
       \vspace{1 em}
	\begin{tabular}{c|c }
		\hline \hline
		   \multicolumn{2}{c}{{Simulated test data}}	\\
		 \hline 
		  Signal & Speech signals from TIMIT \\
		  \hline
		  Room size  & Room 1: ($7 \times 6$) m , Room 2: ($8 \times 8$) m \\
		  \hline
		  Array positions in room & 1 arbitrary position in each room \\
          \hline
		  Source-array distance& 1.5 m  for both rooms\\		  
		  \hline
		  RT$_{60}$ & Room 1: 0.45 s, Room 2: 0.53 s \\
		  \hline
		  SNR & 2 categories: 5 dB, and 15 dB \\		
		  \hline \hline
	\end{tabular}
	\caption{Configuration for generating test data for experiments presented in Section \ref{ssec:DAC} and \ref{ssec:Perturbation}. All rooms are 3 m high.} 
\label{tab:Test}\vspace{0.1em}
\end{table}
\subsection{CNN training}
\label{ssec:Training}
\begin{table}[t]
\footnotesize
\centering
    \setlength{\tabcolsep}{1.7mm}
       \vspace{1 em}
	\begin{tabular}{c c | c | c }
		\hline \hline
		 & SNR = 0 dB & SNR = 10 dB & SNR = 20 dB \\
		 \hline         
		 CNN & $62.3$ & $75.1$ & $90.8$ \\
		 SRP-PHAT & $19.1$& $37.6$ & $43.2$  \\	
		\hline \hline
	\end{tabular}
	\caption{Frame level accuracy ($\%$) for different levels of spatially white noise in matched acoustic condition.}
\label{tab:eval1}\vspace{0.1em}
\end{table}
For the experimental evaluations presented in Sections \ref{ssec:Noise}, \ref{ssec:DAC}, and \ref{ssec:Perturbation}, we consider a ULA with $M = 4$ microphones with inter-microphone distance of 3 cm, and the input signals are transformed to the STFT domain using a DFT length of 256, with $50\%$ overlap, resulting in $K = 129$. To form the classes, we discretize the whole DOA range of a ULA with a $5^{\circ}$ resolution to get $I = 37$ DOA classes. The room impulse responses (RIRs) required to simulate different acoustic conditions are generated using the RIR generator \cite{RIRGenerator}. 

The configuration for generating the training data is given in Table~\ref{tab:Train}. In the training data synthesis, spectrally white noise signals of different levels were convolved with the simulated RIRs of the array. Then, spatially uncorrelated Gaussian noise was added to the training data with randomly chosen noise levels between 0 and 20 dB. In total, the training data consisted of around 5.6 million time frames for the 37 different DOA classes. We used cross-entropy as the loss function and the CNN was trained using the Adam  gradient-based optimizer \cite{Kingma2014}, with mini-batches of 512 time frames. During training, at the end of the three convolution layers and after each fully connected layer, a dropout procedure \cite{Srivastava2014} with a rate of 0.5 was used to avoid overfitting.  
\subsection{Generalization to speech and robustness to noise}
\label{ssec:Noise}
First, we evaluate the ability of the proposed method to localize speech sources, in the presence of additive white noise, in acoustic conditions matching the training scenario. To generate the test data for this experiment, from the training configurations described in Table~\ref{tab:Train}, we chose one of the array positions with 2 m source-array distance in the room denoted as R1. The RIR corresponding to this setup was convolved with 500 different speech samples, each of length 4 s, from the TIMIT database. For different levels of spatially white Gaussian noise, the frame level accuracy of the two methods is given in Table~\ref{tab:eval1}. From the results, it can be seen that the noise trained CNN is able to generalize to speech signals. It also provides a much higher frame level accuracy compared to SRP-PHAT, which suffers from degradation in performance due to noise. 
\begin{table}[t]
\footnotesize
\centering
    \setlength{\tabcolsep}{1.5mm}
       \vspace{1 em}
	\begin{tabular}{c  c | c | c | c}
		\hline \hline
		
	     & \multicolumn{2}{c}{{Room 1}} &  \multicolumn{2}{c}{{Room 2}} \\
		\hline
		          & 5 dB & 15 dB  &  5 dB & 15 dB\\
		 \hline         
		 CNN & 56.2 (57.8)& 69.8 (68.3)& 54.1 (53.6)& 68.2 (68.1)\\
		 SRP-PHAT & 22.6 (17.7)& 33.6 (30.5) & 21.8 (15.1)& 38.4 (33.7)\\
		
		\hline \hline
	\end{tabular}
	\caption{Frame level accuracy ($\%$) for different levels of spatially white noise in different acoustic conditions. Values in brackets show the accuracy when small perturbations in microphone positions are introduced.}
\label{tab:evalMCT}\vspace{0.1em}
\end{table}
\subsection{Different acoustic conditions}
\label{ssec:DAC}
\begin{figure}[t]
\centering
\includegraphics[scale=0.45]{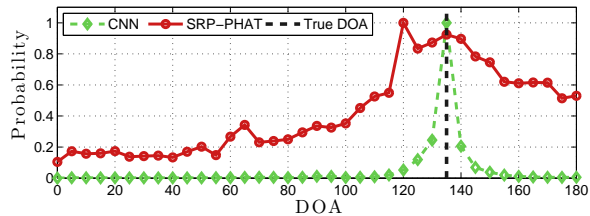}
\caption{DOA probabilities for a speech source positioned at $135^{\circ}$. }
\label{fig:CNNres}
\end{figure}
One of the main challenges for supervised methods for source localization is to adapt to acoustic conditions different from the training conditions. To evaluate this for the proposed method, we generated test data for 2 different acoustic environments with room sizes, reverberation times as well as source-array distance different from the training setup. The details of the configuration for generating the test data is given in Table~\ref{tab:Test}. For each specific room, the same 500 test samples from the previous experiment were convolved with the simulated RIRs. The results for two different SNR levels is provided in Table~\ref{tab:evalMCT}. 

From the results it can be seen that for the unmatched conditions, the proposed method is still able to accurately localize the source for majority of the time frames, however the performance is slightly worse than the matched conditions scenario from the previous experiment. The performance of the proposed method is still considerably better than SRP-PHAT, which fails to provide accurate estimates due to the presence of reverberation and noise. 

An example of the performance of the two methods is depicted in Figure~\ref{fig:CNNres}, which shows the probabilities generated by the two methods for a speech sample, in the test conditions corresponding to Room 1 with SNR = 5 dB (Table~\ref{tab:Test}), where the actual source DOA was $135^{\circ}$. The frame level probabilities were averaged over all active frames and normalized to 1. In this example, it can be seen that the proposed CNN based approach exhibits a clear peak at the true source DOA. In comparison, SRP-PHAT exhibits a much flatter overall distribution, with a false peak at $120^{\circ}$.
\subsection{Robustness to small perturbations in mic positions}
\label{ssec:Perturbation}
In this experiment we investigate the robustness of the proposed method to small perturbations in the microphone positions. The acoustic setup for the test data is the same as in Section~\ref{ssec:DAC}. Small perturbations in the microphone positions were introduced by moving the two middle microphones, in the 4 element ULA, by 5 mm and 3 mm, respectively, in opposite directions along the array axis. The frame level accuracies for this experiment is given in Table~\ref{tab:evalMCT}, values given in brackets.

By comparing the values inside and outside the brackets in Table~\ref{tab:evalMCT}, it can be seen that the CNN based method is more robust to such perturbations compared to SRP-PHAT. A main reason for this is that SRP-PHAT requires exact knowledge of the array geometry for localization whereas for the proposed method, the perturbations lead to local distortions in the input phase map, which the CNN is robust against, due to the weight sharing concept.   
\subsection{Adaptability to real environments}
\label{ssec:Real}
\begin{table}[t]
\footnotesize
\centering
    \setlength{\tabcolsep}{2.5mm}
       \vspace{1 em}
	\begin{tabular}{c  c | c | c | c | c | c}
		\hline \hline
		
	     & \multicolumn{2}{c}{{RT$_{60}$ = 0.160 s}} &  \multicolumn{2}{c}{{RT$_{60}$ = 0.360 s}} & \multicolumn{2}{c}{{RT$_{60}$ = 0.610 s}}\\
		\hline
		          & 1 m & 2 m  & 1 m & 2 m & 1 m & 2 m\\
		 \hline         
		 CNN & 91.8 & 88.7 & 86.8 & 79.4 & 72.3 & 67.3\\
		 SRP-PHAT & 94.4& 69.0 & 87.1& 68.3 & 71.7 & 62.4\\
		
		\hline \hline
	\end{tabular}
	\caption{Frame level accuracy ($\%$) for different distances and reverberation times in real acoustic conditions.}
\label{tab:evalreal}\vspace{0.1em}
\end{table}
Finally, we evaluate the performance of the CNN based method with real data. For this, we used the Multichannel Impulse Response Database from Bar-Ilan university \cite{Hadad2014}. The database consists of measured RIRs with sources placed on a grid of $[0^{\circ}, 180^{\circ}]$, in steps of 15$^{\circ}$, at distances of 1 m and 2 m from the array. For our experiment, we chose the [8, 8, 8, 8, 8, 8, 8] cm array setup \cite{Hadad2014} to get a ULA with $M = 8$ microphones. We trained our CNN for this specific array geometry with simulated data for the R1 setup described in Table~\ref{tab:Train}. The test data was generated by convolving a 15 s long speech segment with the measured RIRs for all the different angles. Spatially white noise was added to the test signal to obtain an average segmental SNR of 30 dB. 

The results for different reverberation times and distances are shown in Table~\ref{tab:evalreal}. From the results, it can be seen that the CNN based approach is able to adapt to real acoustic scenarios even when trained with simulated data and noise signals. When the source is at 2 m, the proposed method clearly outperforms SRP-PHAT. However it can be seen that when the source is closer, SRP-PHAT performs better for lower reverberation times. This can be attributed to the availability of 8 microphones, which improves the spatial selectivity for the SRP based method.

\section{Conclusion}
A CNN based classification method for broadband DOA estimation was proposed that can be trained with noise signals and can generalize to speech sources. Through experimental evaluation, the robustness of the method to noise and small perturbations in microphone positions was shown. The evaluation also demonstrated the ability of the method to localize sources in acoustic conditions that are different from the training data as well as for real acoustic environments. Future work involves testing the proposed approach with different noise types and extending the method for the localization of multiple sound sources.

\balance
\bibliographystyle{IEEEtran}
\bibliography{sapref_WASPAA}
%
%
%
%
%
%
%
%
%

%
%

\end{sloppy}
\end{document}